\documentclass[runningheads,a4paper]{llncs}

\usepackage{amssymb}
\usepackage{graphicx}
\usepackage{wrapfig}

\usepackage{url}
\urldef{\mailsa}\path|{orenled, dcalacci}@media.mit.edu|
\newcommand{\keywords}[1]{\par\addvspace\baselineskip
\noindent\keywordname\enspace\ignorespaces#1}

\usepackage[authormarkup=brackets]{changes}

\usepackage{color}
\definecolor{Orange}{rgb}{1,0.5,0}

\begin{document}

\mainmatter  

\title{Open Badges: A Low-Cost Toolkit for Measuring Team Communication and Dynamics}

\titlerunning{Open Badges}

%
%
\author{Oren Lederman\thanks{corresponding authors}\and Dan Calacci$^*$\and Angus MacMullen\and Daniel C. Fehder\and Fiona E. Murray \and Alex ``Sandy" Pentland}
\authorrunning{Open Badges}

\institute{Massachusetts Institute of Technology,\\
77 Massachusetts Ave, Cambridge, MA 02139, USA\\
\mailsa\\
\url{}}

%
%

\toctitle{Lecture Notes in Computer Science}
\tocauthor{Authors' Instructions}
\maketitle

\begin{abstract}\vspace{-8mm}
We present Open Badges, an open-source framework and toolkit for measuring and shaping face-to-face social interactions using either custom hardware devices or smart phones, and real-time web-based visualizations. 
Open Badges is a modular system that allows researchers to monitor and collect interaction data from people engaged in real-life social settings. In this paper we describe the technical aspects of the Open Badges project and the motivation for its creation.


\end{abstract}

\section{Introduction}
Innovative work is increasingly being done by large, multi-disciplinary teams. As this trend continues, the study of team dynamics - how teams communicate, and how their communication patterns affect their performance - is a growing field of study. 

Prior studies show that team performance can be predicted by social signals, such as turn-taking, response patterns, and conversation balance, as much as conversational content. Pentland found the patterns of communication to be a strong predictor for team performance in the corporate setting~\cite{pentland2012new}. In a study conducted in a European bank call center, his group gathered data on the levels of energy (how much team members contribute to a team), engagement (how team members communicate with each other) and exploration (communication between teams) of each team. Analyzing the results, they found that energy and engagement outside of formal meetings explained one-third of the variation in productivity. In a parallel lab study, Woolley et al. found that groups have a general factor, a collective intelligence, that accounts for 40\% of the variance in performance across a variety of tasks~\cite{woolley2010evidence}. They also showed that the largest factor in predicting group intelligence was the equality of conversational turn taking.

Wearable devices, such as the Sociometric Badges developed in the Human Dynamics group at the MIT Media Lab, help reduce the labor and unconscious bias involved in research studies by automatically quantifying social behavior. These badges also enable researchers to provide real-time feedback to team members. Lowering the cost of this technology and making it more available would allow researchers to deploy it on a large scale.

With these considerations in mind, we have developed a new, inexpensive, open-source platform for measuring face to face group communication using either custom hardware badges or smart phones. In addition, we present a real-time visualization system for team meetings that may be used as an intervention platform in future experiments.

\vspace{-3mm}
\section{Previous Work}
The Human Dynamics group at the MIT Media Lab has been developing wearable devices and mobile frameworks that measure and record face-to-face interaction, location and non-linguistic social signals ~\cite{pentland2005socially,pentland2008honest,pentland2014social}. Starting with the early work on the Sociometer through the more recent Sociometric Badges, the research group has explored the use of sensory data and non-linguistic social signals in order to study human behavior and social interaction.

The first Sociometer, a wearable device designed to measure face-to-face interactions, included an IR transceiver, accelerometers and a microphone~\cite{choudhury2002sociometer}. A shoulder mount placed the Sociometer several inches below the wearer's mouth and reduced the noise created by clothing and movement. The Sociometric Badge~\cite{olguin2010sensor,olguin2011sociometric} presented significant improvements over the Sociometer - longer battery life, larger storage, smaller form factor and better usability. This system evolved into a commercial solution provided by Sociometric Solutions. 

 The group also developed the Meeting Mediator system to quantitatively measure social interaction, and provide real-time feedback to participants in order to facilitate higher group performance and individual satisfaction~\cite{kim2008meeting}.
 
While our work is similar to prior work on the development of the sociometer, ours is the first to present a full platform for face-to-face group interaction research.
The hardware badges we present are significantly smaller and more efficient than previous sensors, and the addition of mobile phones to the research platform may provide researchers with a more widely accessible toolkit for studying group dynamics. 
Ours is also the first to include an open-source analysis and visualization platform in addition to data collection tools.

\vspace{-3mm}
\section{Overview of the Open Badge System}
The Open Badge system consists of three main components: (1) an electronic ``badge" that is worn around the neck and is capable of continuously collecting social interaction data from teams in real-time, (2) a smart phone version of the system, and (3) a modular visualization platform that creates summary visual feedback from the data collected by the badges. 

Prior group interaction studies have marked conversation dynamics as a key factor in predicting team performance. In light of this, our work is mainly focused on speaker detection and the measurement of speaking patterns. These signals provide the most predictive value and cannot be easily implemented through alternative strategies.

The Open Badge system examines the volume of participants' audio output to determine speaking times, disregarding short lower volume events as a rudimentary filter to increase data quality. These data are used to create aggregate statistics such as (i) number of speaking turns per participant, (ii) response patterns between participants, and (iii) frequency of turn-taking. 
Because our system design allows us to reliably record the volumes of each participant independently, we are able to identify simultaneous speaking events between participants, and calculate statistics such as the percentage of overlapped speaking time.

We developed and tested two implementations of electronic badges: (a) custom hardware badges and (b) a parallel implementation for mobile phones. Both types of badges are worn around the neck during meetings or throughout an event.

The two implementations are  low-cost and easy to deploy, with each solution optimized for different technical skills and different settings. We envision the custom hardware badges being deployed in large 1-2 day events, such as conferences and hackathons, where their low price and long battery life give them an advantage over other solutions. These badges requires more preparation and engineering background to fabricate and use.

The mobile application can be installed on any phone running the Android operating system, and therefore requires only minimal technical skills to use. Considering the price of the dedicated mobile device required for running the application, we believe that this solution is more suitable for studies of small numbers of participants at a time.

The following sections describe these two implementations and the visual feedback we developed in greater detail.

\vspace{-3mm}
\subsection{Custom Hardware Badges}

\begin{figure}
    \vspace{-4mm}
	\centering
	\includegraphics[width=0.82\linewidth]{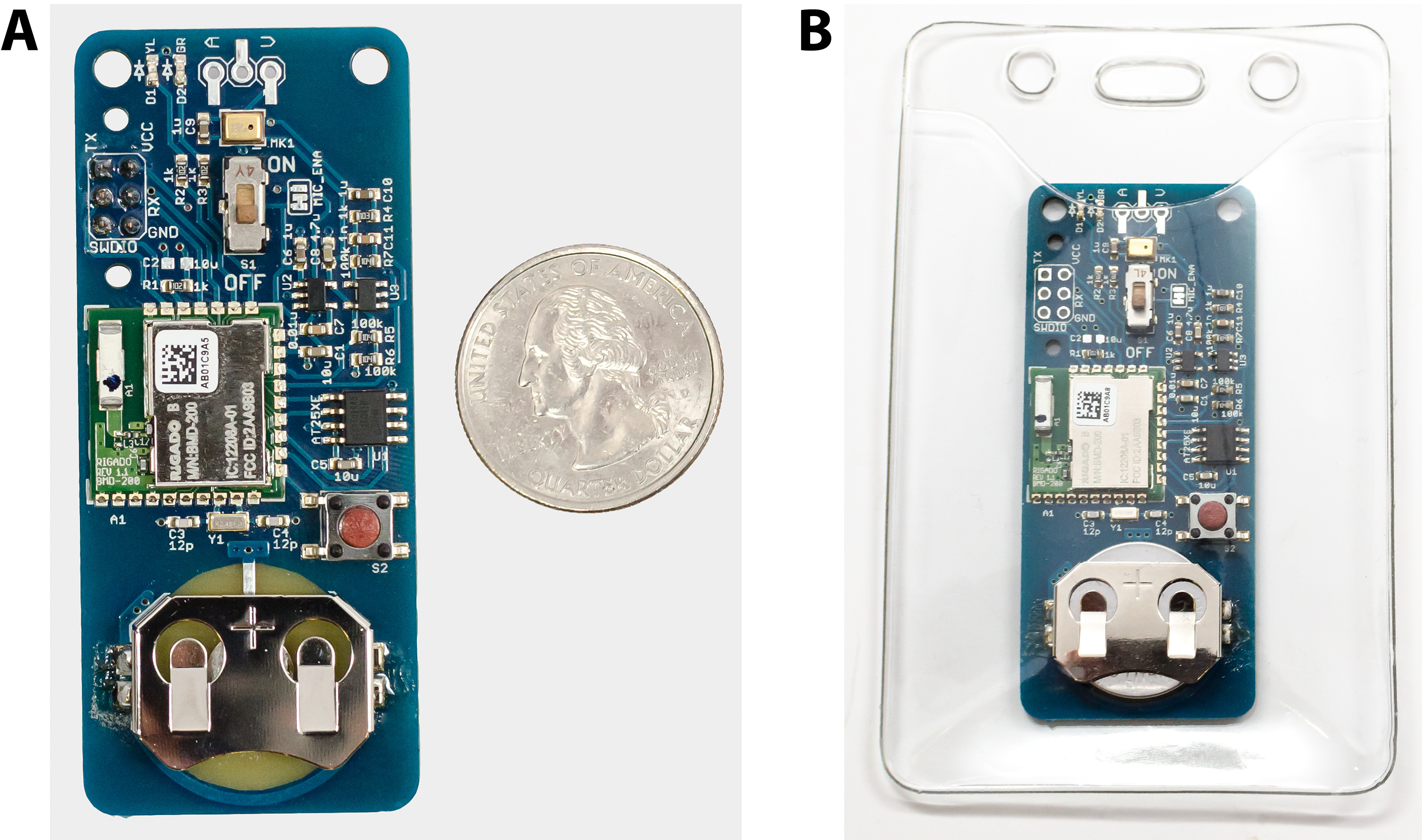}
	\caption{Panel \textsf{(A)} shows the custom hardware badge placed next to a quarter for scale. Panel \textsf{(B)} shows the badge placed in a plastic name tag holder}
	\label{fig:mBadgeSystem}\vspace{-4mm}

\end{figure}

The purpose of the hardware platform is to provide a simple, light, and low-cost electronic badge that can be deployed in large numbers. We achieve this goal by creating a minimalistic circuit design that includes only a small number of components, and that can be hand soldered with a heat gun. The cost per unit\footnote{\small Estimated cost takes into account materials, PCB and assembly} is about \$25, a significant improvement over prior designs. Figure~\ref{fig:mBadgeSystem} shows a prototype of the badge.

The core of the device is the nRF51822 -- a 16MHz ARM Cortex-M0 micro-controller with integrated Bluetooth Low Energy (BLE) that runs our code and transmits the data to a base-station for further analysis. We also use an analog MEMS microphone, an amplifier, a low-pass filter, flash memory, a voltage regulator, a coin cell battery holder, a power switch, a button and several capacitors, resistors and LEDs. These components are mounted on a 3cm x 7cm two-layer PCB small enough to fit in a standard plastic name tag holder.

The badge samples the audio signal in order to detect speaking activity and records proximity to other badges in order to reconstruct the social network of participants and the composition of teams. The microphone signal is sampled at $f_s=8000Hz$ and  averaged over 250 milliseconds. Each badge also performs BLE scans and records RSSI values from other devices every 60 seconds to determine proximity to other badges. The sampling  intervals are programmable.

The samples are stored in flash memory which can hold up to eight hours of data before transmitting them to a base-station. Thanks to its low-power design, the badge can operate for about 40 hours using a single CR2032 coin cell battery.

A base-station is required for monitoring the badges, synchronizing date and time, and periodically pulling data from the badges. Its also responsible for processing the data and sending information to the visualization platform. In our experiments, we used two base station implementations - (1) a Raspberry Pi with Bluetooth and wiFi modules, and (2) an Android smart phone with a dedicated application.


\vspace{-2mm}
\subsection{Mobile App Badges}
In addition to a custom hardware badge, we have developed a mobile phone application that functions as a sociometric badge. 
The goal of this system is to reduce the cost and logistical requirements of deploying team communication experiments.

The mobile app badges simply require distribution of BLE (Bluetooth Low Energy) beacons to participants and for participants to install a mobile application on their phones. Alternatively, the app can be installed on dedicated phones that will be used for a study.
The current version of the mobile phone application is implemented using the Android mobile operating system, the predominant mobile phone platform.

In a mobile app deployment, participants wear the phone around their necks, and the application measures their speaking activity as well as their proximity to other phone badges using BLE scanning.
The mobile application leverages the open-source FUNF\footnote{http://funf.org/} sensing platform for Android phones to collect speaking and BLE scan data while not significantly harming phone battery life.
The mobile application collects volume data from the phone's microphone every 10 milliseconds, and scans for other Low Energy Bluetooth devices every 10 seconds (sampling and scanning intervals are programmable). These data are cached to the phone's local storage, and are sent to a server when a network connection is available.

Before being saved in the database, a band pass filter is applied to the audio data to filter frequencies not in the human vocal range~\cite{olguin2011sociometric}. 
Volume data is then computed from these filtered data. These volume data are smoothed using a rolling median algorithm to reduce noise, and data points are then compared to an experimentally defined threshold to determine when a participant may be speaking.
Speaking events are also determined from the volume modulation produced by each participants' audio signal.
This results in a series of speaking events computed for each participant that consist of a start and end time. 

Bluetooth LE advertisement packages are used for determining proximity. The Android operating system only supports Bluetooth LE advertisements in its latest versions. In order to provide support for more commonly used versions of Android OS, we use external BLE beacons instead. Each beacon periodically broadcasts a unique identifier that is picked up by the BLE scan.

The BLE scan data collected from each mobile phone is filtered to ensure we only collect scans of phones also registered in the participant's group, and scans are then saved in the database. 
These scans search for other registered BLE beacons, and the RSSI value is translated into a rough approximation of physical distance between participants.
The BLE scan data and volume data are then made available through a RESTful API for analysis.

\vspace{-3mm}
\subsection{Visualization Platform}

The visualization platform leverages the API exposed by the database to provide summary visualizations of a group's interaction. 
The platform may be used to generate real-time feedback for participants, or to explore historical meeting data by selecting a past time range to play back. 

\begin{figure}
	\centering
	\includegraphics[width=1.0\linewidth]{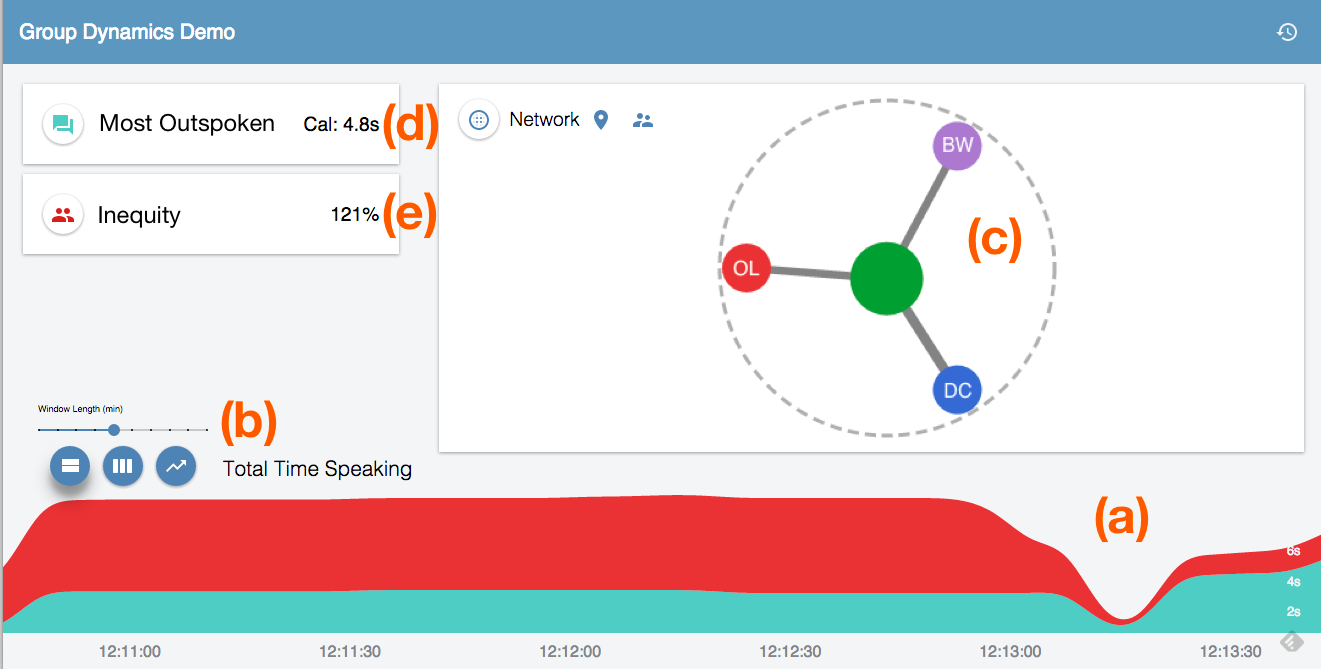}
	\caption{An example visualization that may be built with the web-based platform. The bottom of the screen shows a stacked time-series chart (a) that displays the total speaking time for all participants in the selected time window (b). The main visualization component, (c), is a visualization based on the meeting mediator platform\cite{kim2008meeting}. The two boxes on the left side of the visualization show simple statistics. The first displays the most outspoken participant by name (d), and the second provides a measure of inequality in speaking time for the current time window (e).}
	\label{fig:vizplatform}\vspace{-3mm}
\end{figure}

Figure \ref{fig:vizplatform} shows a screen-shot of the web-based visualization platform. 
It is implemented as a single-page web application that provides feedback using a variety of user-configurable visualizations, and a set of statistics made available to the user.
Figure \ref{fig:vizplatform} shows one example of a visualization configuration that utilizes the platform.
This particular configuration includes a temporal view displayed on the bottom of the screen that shows time-series visualizations of participant data.
The left side of this screen shows real-time meeting statistics.

The network visualization in the top right of the screen is inspired by Kim \textit{et al.}'s work on the Meeting mediator platform, where it was found that this kind of visualization may increase group performance, trust, and engagement, while not increasing cognitive load~\cite{kim2008meeting}.
It displays each participant as a color-coded node on the edge of a network, with a ball in the center.
The edge between a node and the ball in the center grows thicker according to the number of turns that participant has taken in a user-defined time window.
As the participants take more turns as a group, the color of the green ball in the center becomes more intense, with less engaged groups resulting in a pale green ball.
The ball also moves in accordance with the equality of participation in the conversation.
The more equal contributions are between participants, the closer to the center of the network the ball stays.
If a particular participant begins to dominate conversation, the ball will move towards their node over time.

\vspace{-3mm}
\section{Conclusion}

Prior studies have shown that team performance and group intelligence can be predicted by analyzing a group's communication patterns, and that electronic badges can greatly reduce both the cost of such studies and the unconscious bias inherent in recording group interaction data. Feedback created by the analysis of group interaction data has also been shown to help increase group performance, trust, and engagement.

In the interest of extending and enhancing this line of research, we present the Open Badges framework, an inexpensive toolkit for measuring and shaping face-to-face social interactions. We hope that by using these tools, researchers may more easily study the nature of teams, group behavior, and the effects of novel visualizations and interventions on group performance in new settings and at larger scales.

The Open Badge platform is a first step towards creating an open research tool for social interaction that simplifies the technology required for group interaction studies. To this end, we are making the technology available to the wider scientific community. The source code for the Open Badge system and up-to-date information about Open Badge can be found on the Open Badge website.\footnote{http://openbadge.mit.edu}

\bibliographystyle{splncs03}
\vspace{-3mm}
\bibliography{openbadge}

\end{document}